\documentclass[pra,twocolumn,aps, nobibnotes,groupedaddress,showpacs,nofootinbib,floatfix]{revtex4-1}
\usepackage{amsmath, amsthm, amssymb}
\usepackage{array}
\usepackage{amscd}
\usepackage{setspace}
\usepackage{graphicx}
\usepackage{subfigure}
\usepackage{color}
\usepackage{dcolumn}
\newcommand{\bra}[1]{\langle #1|}
\newcommand{\ket}[1]{|#1\rangle}

\begin{document}

\title{Disappearance of entanglement: a topological point of view}
\author{Dong Zhou and Robert Joynt}
\date{\today}
\affiliation{Physics Department, University of Wisconsin-Madison, Madison, Wisconsin 53706, USA}

\begin{abstract}
We give a topological classification of the evolution of entanglement, 
particularly the different ways the entanglement can disappear as a function
of time.
Four categories exhaust all possibilities given the initial quantum state is 
entangled and the final one is not.
Exponential decay of entanglement, entanglement sudden death and sudden birth
can all be understood and visualized in the associated geometrical picture - the
polarization vector representation.
The entanglement evolution categories of any model are determined by the
topology of the state space and the dynamical subspace, the limiting state and 
the memory effect of the environment.
Transitions between these types of behaviors
as a function of physical parameters are also possible. These transitions are 
thus of topological nature.  
The symmetry of the system is also important, since it determines the dimension of the dynamical subspace. 
We illustrate the general concepts with a visualizable model for two qubits, 
and give results for extensions to $N$-qubit GHZ states and W states.
\end{abstract}

\maketitle

\section{introduction}

Quantum entanglement is widely accepted as a useful resource for communication 
and computation \cite{Horodecki09}.
The preservation of this quantity through time is an important goal in 
implementations of information transfer schemes and quantum computers.
In most instances, entanglement is expected to decrease at long times because 
of the inevitable slow leakage of quantum information to the environment. 
However, recent work on models of decoherence of two entangled qubits has shown 
that the manner of this decay can be somewhat surprising. 
    Besides the exponential decay of entanglement, i.e., the half-life (HL) 
behavior, it has been discovered that entanglement as 
a global property may abruptly terminate in a finite time, a phenomenon called entanglement sudden death (ESD) \cite{Yu09review}.
In subsequent work, ESD has been shown to be a rather general phenomenon: 
it can occur when the environment is quantum, classical, Markovian, and 
non-Markovian \cite{Zyczkowski01,Yu04PRL,Yu06PRL,*Yu10Opt,*Yu06Opt, 
Bellomo07,*Bellomo08,Dajka08,*Testolin09,Braun02,*Ficek06,*Mazzola09,Zhou10QIP,
Cole10}.
Oscillatory behavior of the entanglement as a function of time is observed in 
model calculations; this can take the form of entanglement sudden birth (ESB) 
if the two qubits are subject to a common bath \cite{Braun02,*Ficek06,*Mazzola09}.
Other non-monotonic evolutions of entanglement are also possible \cite{Jarvis09,De11}.
Very recently, the existence of ESD has been experimentally confirmed in optical 
and atomic systems \cite{Almeida07,*Laurat07,*Xu10}.

The elements of the density matrix are 
usually analytic functions of time, and the most typical behavior for them 
(or their envelopes) at long times is exponential decay. In ESD, in contrast, the 
entanglement measure  goes to 
zero in a non-analytic fashion; this is because the typical 
entanglement measures are non-analytic functions of the elements of the density
 matrix.  To date, we only have a ``phenomenology" of possible 
behaviors of entanglement. The aim of this paper is to give a soundly based 
theoretical picture. 
    In Sec. \ref{sec:2qubits},
we first categorize qualitatively the various possible 
time evolutions of the entanglement of two qubits and then show how the 
existence of these categories follows from the topology of the state space and 
of the spaces associated with dynamical evolutions, entanglement, and separability.  

For simplicity, we shall present formulas appropriate for the two-qubit case.  
However, the basic results generalize to N qubits.  
    In Sec. \ref{sec:N}, we illustrate the geometrical and topological arguments
    with a pure dephasing model for $N$-qubit GHZ and W states. Finally, we 
    summarize the results in Sec. \ref{sec:summ}.

\section{two-qubit case}
\label{sec:2qubits}

\subsection{state space and entanglement categories}

It is important to choose an appropriate representation of the state space.  
We use the polarization vector representation 
\cite{Kimura03,*Byrd03,Mahler98,*Alicki_Lendi,Joynt_2009}, where the
 density matrix $\rho$, {in the two-qubit case}, is written as 
\begin{align}
\rho =\frac{1}{4}I\otimes I+\frac{1}{4}\sum_{\scriptstyle i,j=I,X,Y,Z \atop \scriptstyle (i,j)\neq(I,I)}
n_{ij}~\sigma _{i}\otimes \sigma _{j}.
\end{align}
Here the components of the real $15$-dimensional vector $\vec n$ are the expectation 
values of all physical observables, i.e., $\vec{n}=\left[ \left\langle I\otimes
X\right\rangle ,\left\langle I\otimes Y\right\rangle,
\ldots\left\langle Y\otimes Z\right\rangle,\left\langle
Z\otimes Z\right\rangle \right]$ where $I=\sigma _{I}$ is the $2\times 2$
identity matrix, and $X=\sigma _{X}$, $Y=\sigma _{Y}$, $Z=\sigma _{Z}$ are the
Pauli matrices. $\vec n$ may be thought of as a generalization of the usual 
Bloch vector. This representation is becoming increasingly popular to describe 
the results of experiments on multiple-qubit systems \cite{Chow10,DiCarlo10}.
For $N$ qubits, the vector has $\dim M = 4^N-1$ components (the 
    number of elements of the $su(2^N)$ algebra).

The state space $M$ is the set of all physically admissible $\vec n$, i.e., those 
that correspond to positive $\rho$. $M$ is a compact, convex manifold with a 
boundary $B_M$. 
The fifteen matrices 
$I\otimes X$, etc., being generators of $SU(4)$ group, form an orthonormal 
basis for the vector space in which $M$ is embedded and the inner product is 
chosen as $\left<A,B\right> =\text{Tr~}AB/4$. 
The metric on $M$ is the one that is induced by this inner product.   

Our measure of two-qubit entanglement is the Wootters' concurrence, denoted by 
$C$ \cite{Wootters98}.  
$C$ is a continuous function on $M$ that satisfies $0\le C\le 1$: $0$ for separable
states and $1$ for maximally entangled states.  
The separable states defined by $C(\vec n)=0$ form a set $S$ that will play an 
important role below.  $S$ is a real manifold with a boundary $B_S$. 
$S\subset M$ and has ``finite volume" in $M$, i.e., $S$ is also $15$-dimensional. 
Like $M$, $S$ is compact and convex.  
{Some points of the boundaries coincide: $B_M \cap B_S$ is not empty and it forms a 
$4$-dimensional manifold. } 
$S$ contains the origin $\vec{n} = \mathbf{0}$, the totally mixed state. Importantly, 
it is known that $S$ contains a ball of radius $1/\sqrt{3}$ centered at the origin 
\cite{Zyczkowski98}.  This is a lower bound for the radius $r_S$ of the maximum 
inscribed ball of separable states, and incidentally gives a connection 
between entanglement and purity. 
The chief difficulty in generalizing entanglement calculations to $N$-qubit 
systems is that there is some degree of arbitrariness in all existing 
definitions of entanglement measure for $N>2$.  
However, there is no arbitrariness in the definition of
 separability: for any $N$, a separable state is still any convex combination 
of products of density matrices for the individual qubits, and any definition 
of entanglement must give zero on these states and no others.  This is all that
 is required for our classification scheme.  It is also still true for $N$
 qubits that $M$ and $S$ are compact and convex, and that $\dim M = \dim S$.  
The shape of $M$ is complicated for large $N$ \cite{Byrd03}, but all pure 
states lie on its surface.  The dimensionality of the submanifold of pure 
states is $2^{N+1}-2$ and that of the pure separable states is $2N$.

The evolution of a quantum system is a smooth curve $\vec{n}(t)$ in $M$. 
This curve induces the continuous function $C(t)=C(\vec{n}(t))$ that is the subject 
of this work.   We shall take $t\in[0,\infty)$ and only consider those trajectories 
with $C(0)>0$ and $C(\infty)=0$, i.e., those that begin in an entangled state and 
end in a separable state. (More general curves are certainly possible, and can also 
be usefully classified by the methods in this paper. For example, it is possible to give criteria for entanglement generation using our scheme.) 
We shall also assume the continuity as a function of time of all components of 
$n(t)$, and all first derivatives. 
Let us denote the set of times when the entanglement is $0$ by $T_0\equiv\{t|C(t)=0\}$.
We define ESD (HL) behavior as any evolution such that $T_0$ is of 
finite (zero) measure.

\begin{figure}[!tbp]
\centering\includegraphics*[width=.8\linewidth]{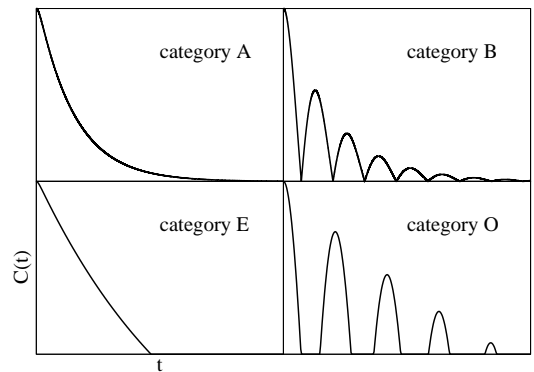}
\caption{Four categories of entanglement evolution.
$\mathcal A$: approaching. $\mathcal B$: bouncing. 
$\mathcal E$: entering. $\mathcal O$: oscillating.
{$\mathcal A$ and $\mathcal B$ belong to the HL class while $\mathcal E$ 
and $\mathcal O$ belong to the ESD class.
}
}
\label{fig:cat}
\end{figure}

Within these two larger classes we must also distinguish
subclasses because of the possibility of oscillations. We define four
categories as shown in Fig. \ref{fig:cat}.  
Category $\mathcal A$ (approaching behavior) is defined by 
$C(t)>0$ for any finite $t$ so that the curve of 
$C\left( t\right) $ never actually hits the horizontal axis. $T_0=\emptyset$ in 
this case. This category includes both monotonic and non-monotonic decay of 
entanglement \cite{De11}. This is the ``canonical" HL behavior.
Category $\mathcal B$ (bouncing behavior) is defined by 
$C(t)=0$ only at isolated times for finite $t$ so that $T_0$ consists of 
isolated points.
Entanglement never quite dies for good in category $\mathcal B$. 
Examples of entanglement evolutions in category $\mathcal B$ can be found in 
Tavis-Cummings model systems when the initial state is in the ``basin of attraction"
\cite{Jarvis09}.
Category $\mathcal E$ (entering behavior) is defined by 
$C\left( t\right) >0$ for $t<t_{d}<\infty$ and $C(t)=0$ for $t\geq t_{d}$. 
 This is the ``canonical" ESD behavior. 
In category $\mathcal O$ (oscillating behavior), $T_0$ consists of disconnected 
intervals of finite width.
Note these four possibilities are exhaustive given that the initial (final) state is
entangled (separable). The division into four categories is a result of two dichotomies: whether $T_0$
is of finite measure and whether $T_0$ is connected. Thus our classification is of topological 
nature.

The existence of these categories can be explained by considering the trajectories
$\vec{n}(t)$ and their relationships to $S$. 
Our basic premise is that three characteristics of any model determine the entanglement
evolution categories: the dynamical subspace $D$ of the model, its limiting point
$\vec{n}_\infty=\lim_{t\rightarrow\infty}\vec{n}(t)$  and the memory effect of the 
environment.  $D$ is defined as the collection of possible trajectories in a model.
It is necessary to introduce $D$ to understand how transitions between categories
can happen. $D$ can be of lower
dimension than $M$ if the evolution has any symmetries or if we limit the set of 
initial conditions in some way. The most important characterstic of any model is
the set $D\cap S$. If $\dim D\cap S=\dim D$, all four categories are 
possible. On the other hand, if $\dim D\cap S<\dim D$, only $\mathcal A$- and 
$\mathcal B$-type behaviors are possible. 
This is because the measure of $T_0$ in $[0,\infty)$ can be no greater than the measure
of $D\cap S$ in $D$.

The limiting state $\vec{n}_\infty$ 
exists for most physical decoherence processes; limit cycles and the like cannot 
be ruled out in general \cite{Terra_Cunha07,*Drumond09},
but we will limit ourselves to the cases where $\vec{n}_\infty \in S$ and is unique.  
This constraint still leaves us with two possibilities:  $\vec{n}_\infty\in \text{Int}(S)$
 (the interior of $S$) and $\vec{n}_\infty\in B_S$. The first case guarantees the 
occurrence of categories $\mathcal E$ and $\mathcal O$ while the second one could 
give rise to all four possibilities depending on other details of $\vec{n}(t)$,
for example, whether $\vec{n}_\infty$ is approached from $S$ or $M\setminus S$.
We take as a working definition that Markovian trajectories satisfy the 
semigroup condition for all possible time partitioning 
\cite{Rivas10,*Breuer09, BreuerBook}. 
Then Markovian evolutions are either in 
categories $\mathcal A$ or $\mathcal E$ while non-Markovian ones are typically in 
categories $\mathcal B$ and $\mathcal O$.

\subsection{case study: a classical noise model}

The aforementioned three characteristics of a model do not uniquely determine the category
of entanglement evolution. Transitions between categories are thus possible by 
tuning some physical parameters of the model \cite{Yu04PRL}.
 We now illustrate this with a general 
decoherence model that is nevertheless of low 
enough dimension that the topology can be visualized.

The  two-qubit Hamiltonian for the model is
\begin{equation*}
H=-\frac{1}{2}\sum_{K=A,B}\left[B_{K} Z_{K} +b_{K}\left(
t\right) \vec{g}_K\cdot \vec{\sigma}_{K}\right],
\end{equation*}
where $\vec{\sigma}_{K}=( X_{K},Y_{K},Z_{K})$.  The fields $
b_{A,B}(t)$ are random time-dependent
fields that decohere qubits A and B and they are not correlated in time. 
$B_K$ is  a static field and $g_K$ is the noise coupling strength.
For simplicity, we take $\vec{g}_A=\vec{g}_B=\vec{g}$.
 An important parameter in the model
is $\theta$, defined by $\cos \theta=\hat{g}\cdot\hat{z}$.  It
is the angle between the energy axis and the noise axis.  $\theta =0$
means that the noise is pure dephasing noise, {while $\theta =\pi /2$ is 
transverse noise. } If the correlation times of $b_{A,B}(t)$
 are short (long) compared with $1/g\cos\theta$, the system is
typically Markovian (non-Markovian) \cite{Zhou10PRA,Galperin06}. 

    Decoherence in this classical noise model comes from the average over all
    noise histories $b_K(t)$. For more detailed discussions, see Ref. 
    \cite{Joynt_2009} {and also the section on Bloch-Wangsness-Redfield theory 
    in Ref. \cite{Slichter96}.

We choose an initial state such that only 
$n_{XX}$, $n_{XY}$, $n_{YX}$, $n_{YY}$, and $n_{ZZ}$ are
nonzero, and this condition is preserved  in the subsequent motion.
We further require $n_{XX}=n_{YY}$ 
and $n_{XY}=-n_{YX}$, which leaves only three independent parameters.
This defines the dynamical subspace $D$ as a $3$-dimensional slice 
of $M$.  
We refer to this as $D_{3}$.  
$D_3$ is large enough to accommodate essentially any
decoherence dynamics given that (1) the two qubits are non-interacting; (2)
noises on the two qubits are uncorrelated in time; 
(3) the effect of dephasing and relaxation can be separated; 
(4) the initial state is in $D_3$. 
 Thus $D_{3}$ is the dynamical manifold of a rather general class of decoherence 
 processes.  Note that $n_{XX}^{2}+n_{XY}^{2}=R^{2}$ is conserved in $D_3$.
The positivity of the density matrix requires
\begin{align}
a_2\ge0 &~ \Rightarrow~ 2R^2+n_{ZZ}^2\le3 \label{eq:a2}\\
a_3\ge0 &~ \Rightarrow~ n_{ZZ}\le1-2R^2,~\text{and}~ n_{ZZ}\ge-1 \\
a_4\ge0 &~ \Rightarrow~ 2R+n_{ZZ}\le1 
\end{align}
where $a_i$ are the coefficients of the characteristic polynomial 
det$(xI\otimes I-\rho)=\sum_{j=0}^N(-1)^ja_j x^{N-j}$ \cite{Kimura03}.

Applying these inequalities, we find that $D_3$ is
a cone in the ($n_{XX}$,$n_{XY}$,$n_{ZZ}$) coordinates.  
Its cross-section, as shown in Fig. \ref{fig:contour}, is an isosceles
triangle with height $2$ and base $2$ and the full
manifold is generated by rotation about the $n_{ZZ}$ axis. 
The concurrence is given by
\begin{align}
C= \max\left\{0,R-(1+n_{ZZ})/{2}\right\}.  \label{eq:C}
\end{align}
Separable states $S\cap D_3$ form a set with a spindle shape on top 
and the entangled states form a
torus-like shape on the bottom with triangular cross sections. 
The direction of the gradient of $C$ is indicated by the filled arrows 
in Fig.\ref{fig:contour}
and the maximally entangled Bell states
$\left\vert \Psi \right\rangle =\frac{1}{2
}\left( \left\vert 01\right\rangle +e^{i\phi }\left\vert 10\right\rangle
\right)$  with $C=1$ 
 form the circle $R=1$ and $n_{ZZ}=-1$ on the lower surface of the cone.
{It is tempting to consider the minimum Euclidean distance 
to separable states in the 
polarization vector representation as a geometric measure of entanglement
\cite{Wei03, Cole10}.
It indeed works in $D_3$ but whether it qualifies as an entanglement 
measure in general is still an open question \cite{Verstraete02}.
}

\begin{figure}[tb]
\centering\includegraphics*[width=.8\linewidth]{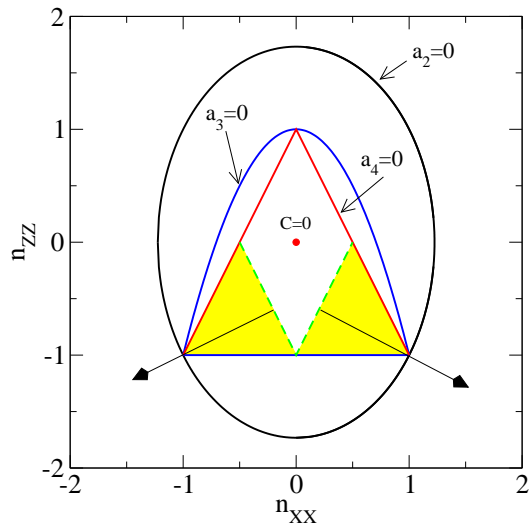}
\caption{(Color online) Cross section of the dynamical subspace $D_3$
with $n_{XY}=0$. 
$a_2=0$ is the ellipse.
$a_3=0$ gives the parabola and the bottom of the isosceles triangle.
$a_4=0$ sets the two sides of the triangle.
The entangled region is shaded where the filled arrows denote increasing direction of the concurrence.
The green dashed line is the boundary of entangled and separable states, i.e.,
$B_S\cap D_3$.
The fully mixed state is denoted by a red dot.
This isosceles triangle corresponds to a projection of the tetrahedron in Ref.
\cite{Horodecki96PRA}. 
}
\label{fig:contour}
\end{figure}

We thus fully characterized the entanglement topology of $D_3$. 
Now we construct time evolutions in $D_3$.
 The initial state is taken as the generalized Werner state $\omega
_{r}^{\Psi }=r\left\vert \Psi \right\rangle \left\langle \Psi \right\vert
+\left( 1-r\right) I_{4}/4$ \cite{Zhou10QIP,Werner89}.  
Without any loss of generality, we choose $\phi=0$ in $\ket{\Psi}$, 
giving $n_{XY}=0$. 
This allows us to visualize the state and entanglement evolutions
in a $2$-dimensional picture.
The initial state is
$\vec{n}(0)=(n_{XX},n_{ZZ}) =(r,-r)$
and the state trajectory is given by
\begin{align}
n_{XX}\left( t\right)  &=\left\langle X\otimes X\right\rangle(t) =r~\zeta
^{AB}(t), \\
n_{ZZ}\left( t\right)  &=-r~e^{-\Gamma_1^{AB}t}.
\end{align}
where $\Gamma_1^{AB}=\Gamma _{1}^{A}+\Gamma_{1}^{B}$ is the overall longitudinal 
relaxation rate. 
$\zeta^{AB}=\zeta^A\zeta ^{B}$ describes overall dephasing process.
Note $\zeta\left( 0\right)=1$ and 
$\zeta \left( \infty \right) =0$ if dephasing occurs. 
The details of $\zeta(t)$ and $\Gamma_1$ are model-dependent.
For Markovian noises, the Bloch-Wangsness-Redfield theory applies and we have 
$\zeta^K(t)=\exp(-\Gamma_2^K t)$, where 
$\Gamma_2^K = \Gamma_1^K/2 + \Gamma_\phi^K$,
$\Gamma_1^K=g_K^2\sin^2\theta S^K(B_k)/2$ and  
$\Gamma_\phi^K = g_K^2\cos^2\theta S^K(0)/2$ for $K=A,B$ \cite{Slichter96}. 
Here $S^K(\omega)$ is the power spectrum function (Fourier transform of the 
noise autocorrelation function) of the classical noise process $b_K(t)$.
For non-Markovian noises, non-exponential behaviors in $\zeta(t)$, such as 
damped oscillations, are possible \cite{Zhou10QIP}.

The direction of the trajectory in $D_{3}$ is determined by
the relative weight of dephasing and relaxation noise on the qubits. 
The trajectories are characterized by two parameters, $r$ which fixes the
initial position, and $\theta$, that weights the two types of noise.  In
the case of pure dephasing ($\theta=0$), $n_{ZZ}$ is constant and the trajectory is
horizontal with $\vec{n}_{\infty }=(0,-r)$. 
{This is a visualization of decoherence free subspace 
where the constants of the decoherence dynamics can be used to encode information
\cite{Lidar98PRL,*Bacon00}. }
By contrast, when there is relaxation noise $(\theta\neq0)$, the trajectory also moves
vertically, so that when both types of noises are present, the trajectory will 
 have $\vec{n}_{\infty }=\left( 0,0\right)\in \text{Int}(S)$.   Only
one trajectory with $r=1$ and $\theta=0$ (pure dephasing), 
gives HL behavior since it remains in the entangled manifold except for 
$\vec{n}_{\infty }=(0,-1)\in B_S$.  All other trajectories show ESD.  
We may summarize the situation in topological language by saying that 
{ HL behavior requires that $\vec{n}_{\infty}\in B_{S}$}. 

\begin{figure*}[t!]
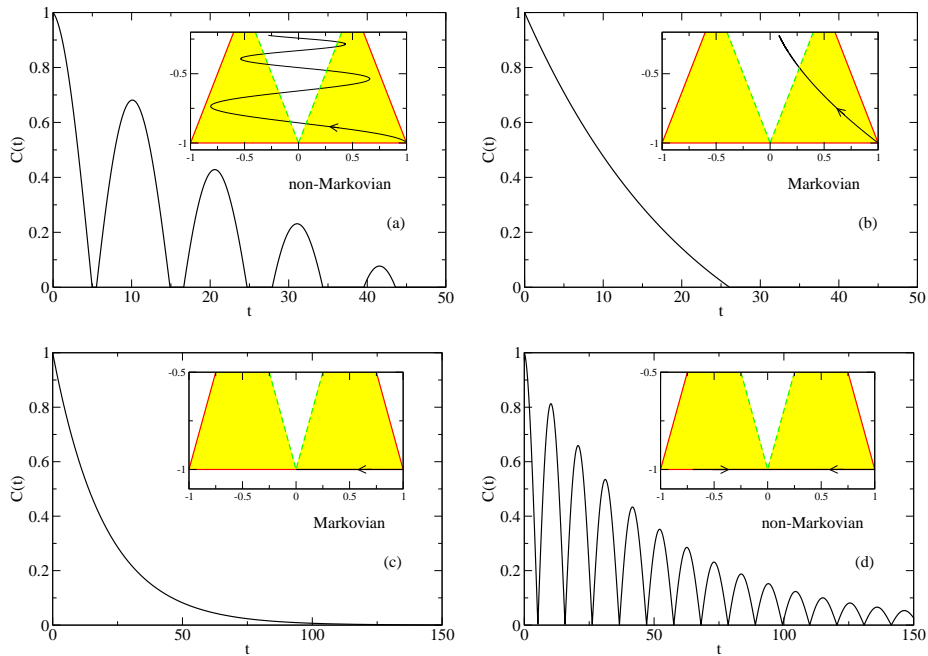

\centering
\subfigure{
\includegraphics*[width=.33\linewidth]{NM_traj.eps}
}
\subfigure{
\includegraphics*[width=.33\linewidth]{M_traj.eps} } \\
\subfigure{
\includegraphics*[width=.33\linewidth]{M_traj_pd.eps} }
\subfigure{
\includegraphics*[width=.33\linewidth]{NM_traj_pd.eps} }
\caption{(Color online)
Concurrence evolution and state trajectory (inset) in the presence of Markovian and non-Markovian noise.
The arrows denote time direction. 
(a) and (b) have both dephasing and relaxation present,
while (c) and (d) are the pure dephasing case.
The initial state is Bell state $\left|\Psi+\right>$.
We used $\zeta^{AB}=e^{-0.02t}\cos 0.3t$ for non-Markovian noise 
and $\zeta^{AB}=e^{-0.05t}$ for Markovian noise.
For (a) and (b), $\Gamma_1^{AB}=0.03$.
}
\label{fig:dynamics}
\end{figure*}

To pin down the precise category of the entanglement evolution, we note that
\begin{align}
C(t) =\max \left\{ 0,r\left[ \zeta ^{AB}\left( t\right)
-\xi \left( t\right) \right] \right\} 
\end{align}
where $\xi(t) =( 1-re^{-\Gamma _{1}^{AB}t})/2$.  
If the noise is Markovian (non-Markovian) then $\zeta$ is monotonically
decreasing  (oscillatory)
\cite{Bellomo08,Zhou10QIP,Dajka08,Mazzola09}. 
In the latter case we find category $\mathcal O$, as the state
trajectory enters and leaves the separable region as seen in Fig. \ref{fig:dynamics}a.
In the case of Markovian noise, once the state leaves the entangled 
region, it leaves forever and we find category $\mathcal E$, as in 
Fig. \ref{fig:dynamics}b.  There is a transition
between this ESD behavior and HL behavior at the critical point $r=1$ and 
$\theta=0$.  At this point we have $\vec{n}_{\infty}=(0,-1)\in B_S$.
The transition is characterized by the fact that the ``critical" trajectory 
intersects $B_{S}\cap D_3$ but not $\text{Int}(S)\cap D_3$.  Markovian evolution 
yields category $\mathcal A$ behavior, as shown in
Fig. \ref{fig:dynamics}c while non-Markovian behavior yields category 
$\mathcal B$ behavior as shown in Fig. \ref{fig:dynamics}d.  
Another perspective is that the dynamical subspace for these trajectories with
$r=1$, $\theta=0$ is the bottom disc $D_2$ of $D_3$ and 
$\dim D_2\cap S=0<\dim D_2=2$.
The transition of $T_0$ from the ESD class to HL class is thus a result
of an topological transition of the dynamical subspace from $D_3$ to $D_2$.   

\section{$N$-qubit case: GHZ and W states}
\label{sec:N}
Our examples have been drawn from the entanglement evolution of two-qubit 
systems.  This is convenient, since the two-qubit concurrence is easily 
evaluated, and the lower dimensionality makes the examples relatively easy to 
visualize.  However, it should be clear that the precise definition of 
entanglement is not important for the topological categorization.  Only the 
definition of separability, which alone determines the set $S$, is crucial. 
As we saw above, the topological properties of $S$, particularly 
$\dim S = \dim M$, carry over to $N$-qubit systems. 
There are indications that as the dimension increases, the percentage of the separable 
states in the physical states decreases \cite{Zyczkowski98,Zyczkowski99,Gurvits03,*Gurvits05}.

The entanglement measure for general multi-partite quantum system is a topic of 
current research \cite{Miyake03,*Osborne05,*Wong01,*deOliveira09,*Ou07,*Jung09,*Carvalho04,*Mintert05,Dur00,Vidal02}. 
{The only easily computable entanglement measure for arbitrary $N$-qubit mixed 
states is the negativity \cite{Vidal02}, which measures the subsystem entanglement
with respect to bipartite partitioning of the whole system.}
It is defined as 
\begin{align}
N(\rho) = \frac{\|\rho^{T_A}\|_1-1}{2}.
\end{align}
Here the norm is taken to be trace norm and $T_A$ denotes partial transpose on 
one of the two subsystems.
$N(\rho)$ corresponds to the absolute value of the sum of the negative 
eigenvalues of $\rho^{T_A}$, which according to Peres-Horodeck's criterion  
reveals the entanglement in $\rho$ \cite{Peres96,*Horodecki96Peres}.
Here we use $N(\rho)=0$ as a working criteria for separability.

We will use the negativity to illustrate the entanglement evolutions of 
$N$-qubit GHZ and W states due to pure dephasing noise.
The GHZ state and W states are of interest to the quantum computing community 
since they possess different types of multi-partite entanglement and are used 
in various protocols such as quantum secret sharing, teleportation and super 
dense coding \cite{D'Hondt05,Dur00,Verstraete03,Tittel01,Yeo06}.
The $N$-qubit GHZ and W states are defined to be 
\begin{align}
\left|{\text {GHZ}}\right> =& \frac{\left|0\right>^{\otimes N}+ \left|1\right>^{\otimes N}}{\sqrt{2} }\\
\left|{\text {W}}\right> =& \frac{1}{\sqrt{N}}\left(\left|0\ldots01\right> + \left|0\ldots10\right>+\cdots+\left|1\ldots00\right>\right).
\end{align}
The dephasing process on a single qubit can be described in terms of two Kraus operators \cite{Nielsen00Chuang, Weinstein09jan}
\begin{align}
E_0=\begin{bmatrix}
1 &  0 \\
0 & \zeta(t)\end{bmatrix}, \qquad
E_1=\begin{bmatrix}
0 &  0 \\
0 & \sqrt{1-\zeta^2(t)}\end{bmatrix}.
\end{align}
Note $\zeta(t)$ is smooth and characterize the dephasing process  in  the $xy$ 
plane in the Bloch vector representation, just as in $D_3$.
For simplicity, we assume that all qubits are independent and they share the 
same dephasing function $\zeta(t)$.
Thus the full dynamics of the $N$-qubit system can be described by 
\begin{align}
\rho(t) = \sum_{i_1,i_2,\ldots,i_N}E_{i_1}E_{i_2}\cdots E_{i_N} \rho_0 E_{i_N}^\dagger \cdots E_{i_2}^\dagger E_{i_2}^\dagger.
\end{align}

For the negativity, we make the $(1)(N-1)$ bipartite partition, i.e., the partial
transpose is applied to one of the $N$ qubits. Due to the permutation 
symmetry  of GHZ and W states, as well as the $N$-qubit Kraus operators, 
it does not matter which qubit is picked, indicating that the entanglement is
evenly distributed among the $N$ qubits.
Given these assumptions, the negativity of the  $N$-qubit GHZ  and W states are  
given by 
\begin{align}
N_{\text{GHZ}}=&|\zeta|^3(t)/2\\
N_{\text{W}} = &\frac{\sqrt{N-1}}{N} |\zeta|^2(t).
\end{align}
Notice that due to the smoothness of $\zeta(t)$, entanglement evolutions in 
categories $\mathcal E$ and $\mathcal O$ cannot happen.
This can be explained by the topological argument as follows.

For both the GHZ and W states, the dephasing process does not expand the Hilbert space. 
For the $N$-qubit GHZ states, 
the density matrix can be expanded with identity and elements of the $su(2)$
algebra {
\begin{align}
\rho_{\text{GHZ}} = \frac{1}{2} \left(I + \sum_{i=X,Y,Z} n_i \sigma_i \right),
\end{align}
where $\sigma_i$ are defined on the two cat states, for example, 
$\sigma_X = \ket{0}^{\otimes N}\bra{1}^{\otimes N} + \ket{1}^{\otimes N}\bra{0}^{\otimes N}$.
}
In other words, the dynamical subspace $D_{\text{GHZ}}$ for GHZ states under
dephasing is a $3$-dimensional Bloch ball, as seen in Fig. \ref{fig:bloch2}.
{The negativity is given by 
\begin{align}
N(\rho_{\text{GHZ}}) = \frac{n_X^2 + n_Y^2}{2}.
\end{align}
}
The separable states in terms of negativity live on the vertical symmetry 
axis.
We have 
\begin{align*}
\dim D_{\text{GHZ}}\cap S=1< \dim D_{\text{GHZ}}=3
\end{align*}
 thus only entanglement evolutions in categories $\mathcal A$ and $\mathcal B$ are possible.

\begin{figure}[tbp]
\centering\includegraphics*[width=.8\linewidth]{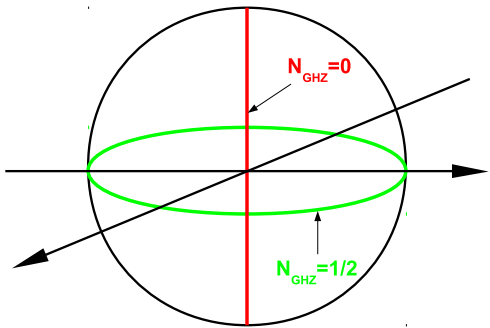}
\caption{(Color online) Effective Bloch sphere representation for the
dynmaical subspace $D_{\text{GHZ}}$ . The north and south pole are either $%
\left|0\right>^N$ or $\left|1\right>^N$.
States resting on the red line connecting north and
south pole have zero negativity. States on the green equator are 
maximally entangled.}
\label{fig:bloch2}
\end{figure}

For the W states, the density matrix is expanded with identity and elements of 
the $su(N)$ algebra. 
Thus the dynamical subspace $D_{\text{W}}$ is $N^2-1$ dimensional.
However, when calculating negativity, the partial transpose introduces $N$ extra 
bases into the Hilbert space and $\rho^{T_A}(t)$ contains elements in $su(2N)$ algebra. 
The partial tranpose moves $N-1$ elements $\beta_i$ and their complex conjugates
$\beta^*$ in the original density matrix to 
the blocks of the newly introduced bases. 
The positions of these elements $\beta_i$ are related to the 
partial transpose: suppose the partial transpose acts on 
the $k$'th qubit, then $\beta_i$ are the coefficients of 
$\left|\ldots 0 \ldots \right>\left< \ldots 1 \ldots \right|$, where $0$ and 
$1$ label the $k$'th qubit.
    Take the $3$-qubit case for example, the component $\beta_1\ket{001}\bra{100}$ is
    mapped into $\beta_1\ket{101}\bra{000}$,  
    where the partial transpose is taken on the first (left-most) qubit.
        On the other hand, components such as $\ket{001}\bra{010}$ or 
        $\ket{100}\bra{100}$ are left unchanged under the action of partial 
        transpose on the first qubit.

The negativity for any density matrix expandable by the W state bases is given by 
\begin{align}
N(\rho_{{\text{W}}}) = \frac{1}{N} \sqrt{\sum_{i=1}^{N-1}|\beta_i|^2}
\end{align}
Note that the condition of separability $N(\rho_{{\text{W}}})=0$ eliminates 
$2(N-1)$ degrees of freedom.  
Thus 
\begin{align*}
\dim(D_{\text{W}}\cap S)=(N-1)^2 < \dim D_{\text{W}} = N^2-1
\end{align*}
 and we conclude 
that entanglement evolutions in categories $\mathcal E$ and $\mathcal O$ are 
not possible.

\section{summary}
\label{sec:summ}

In summary, we found that there are four and only four types of natural 
behaviors for the time evolution of entanglement,
given the initial state is entangled and the final one separable. 
These categories are determined by the dimensionality and intersection 
properties of sets in $M$. Since these properties are preserved by 
continuous deformations, they are topological.
Three characteristics determine the categories: the dynamical subspace 
$D$ of the model, the limiting state $\vec{n}_\infty$ and the memory effect of the environment.
Category $\mathcal A$ occurs (typically) for Markovian systems when 
$\vec{n}_\infty\in B_S$, category $\mathcal B$ occurs (typically) for non-Markovian 
systems when $\vec{n}_\infty\in B_S$. Category $\mathcal E$ occurs (typically) 
for Markovian systems, while category $\mathcal O$ occurs (typically) for non-Markovian 
systems. For those two categories, $\vec{n}_\infty$ can be either on the boundary or 
in the interior of $S$. 

Model studies of entanglement evolution have shown a wide variety of behaviors, and 
a unifying picture of these behaviors has been lacking.  The topological approach 
given here provides such a picture.  The qualitative behavior is determined by 
relative dimensions of the dynamical subspace, the space of separable states, and 
the space formed by their intersection.  The determination of these dimensions, 
together with the location of the asymptotic point, tells us when transitions between
 different types of evolution can be expected.  Precise means of computing the 
dimensions, including the important role of symmetry, will be given in a future 
publication.

\begin{acknowledgments}
We thank J. H. Eberly, S. N. Coppersmith, A. De, A. Lang, G.-W. Chern and R. C. 
Drumond for helpful discussions and correspondence. This work was supported by 
NSF-DMR-0805045, by the DARPA QuEST program, and by ARO and LPS W911NF-08-1-0482. 
\end{acknowledgments}

\bibliography{/home/nos/projects/refs}       

\end{document}